\documentclass[aps,twocolumn,prb,showpacs,preprintnumbers,amsmath,amssymb]{revtex4}
\newcommand{\lsim} 
 {\ \raise.35ex\hbox{$<$}\kern-0.75em\lower.5ex\hbox{$\sim$}\ }
\newcommand{\gsim}
 {\ \raise.35ex\hbox{$>$}\kern-0.75em\lower.5ex\hbox{$\sim$}\ }

\usepackage{graphicx}
\usepackage{dcolumn}
\usepackage{bm}
\usepackage{amsmath}
\usepackage{times}
\usepackage[usenames]{color}

\usepackage{ulem}

\begin{document}
\title{Photoinduced correlated electron dynamics in a two-leg ladder Hubbard system}
\author{Hiroshi~Hashimoto  and Sumio~Ishihara} 
 \affiliation{Department  of Physics,  Tohoku University,  Sendai 980-8578,  Japan} 
  \affiliation{CREST-JST,  Sendai 980-8578,  Japan} 

\date{\today}
\begin{abstract}  
Photoinduced carrier dynamics in a correlated electron system on a coupled two-leg ladder lattice are studied. 
The two-leg ladder Hubbard model is analyzed by utilizing the exact diagonalization method based on the Lanczos algorithm in finite size clusters. 
In order to reveal the transient carrier dynamics after photoirradiation, we calculate the low-energy components of the hole kinetic energy, the pair-field correlation function, the optical conductivity spectra and others. 
It is shown that the photoinduced metallic-like state appears in a half filled Mott insulating state, while the low-energy carrier motion is suppressed by photoirradiation in hole doped metallic states. 
These photoinduced changes in electron dynamics are associated with changes in the carrier-pair coherence, 
and are not attributed to a naive thermalization but to a ladder-lattice effect. 
Based on the numerical results, optical controls of hole pairs by using the double-pulse pumping are demonstrated. 
Implications to the recent optical pump-probe experiments are presented. 
\end{abstract}


\maketitle



%
%

%




\section{Introduction}

Optical manipulations of macroscopic phenomena in solids are widely accepted as greatly attracted subjects in recent condensed matter physics.~\cite{nasu,PIPT_JPSJ} 
Ultrafast controls of magnetism~\cite{miyasaka,ehrke}, ferroelectricity~\cite{uemura,miyamoto}
and electrical conductivity~\cite{ogasawara,iwai,okamoto,okamoto1,okamoto2,okamoto3,tajima,kawakami}, as well as superconductivity~\cite{fausti,nicoletti},
 have tried to be realized in a wide variety of materials by utilizing the recently developed laser pulse technology. 
Strongly correlated electron systems, such as transition-metal oxides, organic molecular solids, and rare-earth compounds, 
are the plausible candidate materials, where a number of degrees of freedom are entangled with each other under strong electron-electron and electron-lattice interactions. 
Observations of transient changes in macroscopic quantities and comprehensions of the microscopic mechanisms are required as problems of great urgency from the fundamental and application viewpoints. 

Low dimensional correlated electron systems are attractive targets for the ultrafast optical controls of electronic states. 
Quasi one- and two-dimensional cuprate oxides~\cite{ogasawara,okamoto1,okamoto2,fausti,nicoletti,lorenzana,iyoda,lenarcic},
halogen-bridged complexes~\cite{iwai, okamoto3}
and BEDT-TTF molecular solids~\cite{tajima, kawakami} are the examples. 
This is attributed to a rich variety of equilibrium states and the simple lattice structures. 
From the theoretical viewpoints,  transient electron dynamics have been examined numerically and analytically 
in the one- and two-dimensional
Hubbard~\cite{yonemitsu,lenarcic2,lu,lu2} and $t-J$~\cite{iyoda,lenarcic} models which are accepted as minimal models for the low-dimensional correlated systems. 

Correlated electrons on a ladder lattice are recognized to be in between the one- and two-dimensional systems. 
The superconductivity observed in the two-leg ladder cuprates, Sr$_{14-x}$Ca$_x$Cu$_{24}$O$_{41}$, suggests  
unique electronic structures on a ladder lattice.  
Insulating states associated with a charge order at $x=0$ is changed into a metallic state by substitution of Sr by Ca, 
and superconductivity appears under high pressure.~\cite{nagata,motoyama,uehara,osafune}
Short-range singlet spin correlation and spin excitation gap have attracted much attentions from the viewpoint of the superconductivity caused by doping of holes into the ladders.~\cite{dagotto_rice} 
A number of theoretical examinations for the equilibrium electric and magnetic structures, as well as doublon--holon dynamics have been performed so far, in the two-leg ladder Hubbard and $t-J$ models.~\cite{noack,noack2,riera,dagotto,tsunetsugu,dias}
Recently, the photoinduced transient states were examined in the insulating ($x=0$) and metallic ($x=10$) two-leg ladder cuprates by utilizing the ultrafast optical pump-probe experiments. ~\cite{fukaya,fukaya1,fukaya2}
The transient optical measurements provide a hint to reveal unique spin and charge dynamics as well as their roles on the superconductivity in the two-leg ladder cuprates. 

In this paper, the photoinduced carrier dynamics in a correlated electron system on a two-leg ladder lattice are studied. 
The two-leg Hubbard model is analyzed by utilizing the exact-diagonalization method base on the Lanczos algorithm in finite size clusters. 
In order to reveal the transient dynamics, we calculate the low-energy components of the kinetic energy, the pair-field correlation function, the optical conductivity spectra and others as functions of time. 
It is found that the low energy carrier dynamics in metallic states are distinguish from those in the insulating states; 
the photoinduced metallic-like state appears at half filled Mott insulating state, while the low-energy carrier motion is suppressed by photoirradiation in hole doped metallic states. 
These photoinduced changes in the dynamics are associated with the changes in pair coherence of carriers. 
Based on the calculated results, the optical controls of hole pairs by utilizing the double-pulse pumping are demonstrated. 
Implications to the recent optical pump-probe experiments are presented. 

In Sect.~\ref{sec:model}, the theoretical model and the numerical calculation method are introduced. 
In Sect.~\ref{sec:kinetics}, numerical results for the photoinduced carrier dynamics are presented. 
In Sect.~\ref{sec:double}, numerical demonstrations of the photo-control carrier dynamics by utilizing the double-pulse pumping are shown. 
Section~\ref{sec:summary} is devoted to discussion and summary. 

\section{Model and Method}
\label{sec:model}

The Hubbard model on a two-leg ladder lattice is defined by 
\begin{align}
{\cal H}=
&- \sum_{\langle ij \rangle \alpha \sigma} \left ( t c_{i \alpha \sigma}^\dagger c_{j \alpha \sigma} + H. c. \right) \nonumber \\
&- \sum_{i \sigma} \left ( t' c_{i 1 \sigma}^\dagger c_{i 2 \sigma} + H.c. \right) 
+U \sum_{i \alpha} n_{i \alpha \uparrow} n_{i \alpha \downarrow} , 
\label{eq:ham}
\end{align}
where $c_{i \alpha \sigma}^\dagger$ and $c_{i \alpha \sigma}$ are the creation and annihilation operators for an electron at the $i$-th rung and the right $(\alpha=1)$ or left $(\alpha=2)$ leg with spin $\sigma(=\uparrow, \downarrow)$, and 
$n_{i \alpha \sigma}=c_{i \alpha \sigma}^\dagger c_{i \alpha \sigma}$ is the number operator. 
The first and second terms represent the electron hoppings along the leg and rung in the two-leg ladder lattice, respectively, 
and the third term describes the on-site Coulomb interaction. 
A symbol $\langle ij \rangle$ represents the nearest-neighbor $ij$ pair along a leg.  
We introduce an anisotropy of the hopping integral as $r_t = t'/t$ which is nearly equal to one in the ladder cuprates. 
The photoinduced dynamics are examined by introducing the pump pulse photons in the first term in Eq.~(\ref{eq:ham}) as 
the Peierls phase given by 
\begin{align}
t \rightarrow t e^{i A(\tau)} , 
\label{eq:peierls}
\end{align}
where a lattice constant, light velocity, and elementary charge are taken to be units, 
and the light polarization is parallel to the leg direction. 
The vector potential of the pump photon pulse at time $\tau$ is taken to be a Gaussian form as 
\begin{align}
A(\tau)=A_p e^{-\tau/(2\gamma_p^2)} \cos \omega_{p} \tau , 
\label{eq:vp}
\end{align}
with amplitude $A_p$, frequency $\omega_{p}$ and pulse width $\gamma_p$. 
A center of the pump photon pulse is located at $\tau=0$. 

Electronic states before and after the photon pumping are calculated by using the exact-diagonalization method based on the Lanczos algorithm. 
We adopt finite-size clusters of $N(=2 \times N/2)$ sites up to $N=14$, where the periodic- and open-boundary conditions are imposed along the leg.  
The time evolutions of the wave function under the time-dependent Hamiltonian are calculated as~\cite{park, prelovsek} 
\begin{align}
| \Psi(\tau+\delta \tau) \rangle & =e^{-i{\cal H}(\tau) \delta \tau } |\Psi(\tau) \rangle
\nonumber \\
& = \sum_i^M e^{-i { \varepsilon}_i \delta \tau} |\phi_i \rangle \langle \phi_i | \Psi(\tau)\rangle , 
\label{eq:wave}
\end{align}
where $|\phi_i \rangle$ and ${ \varepsilon}_i$, respectively, are the eigen-state and the eigen-energy in the order-$M$ Krylov subspace in the Lanczos process, $\delta \tau$ is a time step, and ${\cal H}(\tau)$ is the time-dependent Hamiltonian where the pump pulse is introduced as Eq.~(\ref{eq:peierls}).  
We chose, in most of the numerical calculations, $M=15$ and $\delta \tau=0.01/t$, which are sufficient to obtain the results with high enough accuracy. 

All energy and time parameters in the numerical calculations, respectively, are given as units of $t$ and $1/t$, which are about 0.5eV and 8fs in the ladder cuprates. 
In most of the calculations, we chose $U/t=6$ and $\gamma_{p} t = 5$. 
Hole density measured from the half filling are denoted by $x_h(\equiv 1-N_e/N)$ with the number of electrons $N_e$.

\section{Results}
\label{sec:results}

\begin{figure}[t]
\includegraphics[width=0.8\columnwidth,clip]{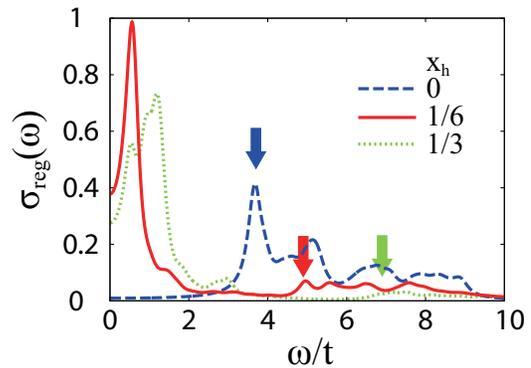}
\caption{
(Color online)
Optical conductivity spectra before photon pumping. 
Dashed, solid and dotted curves represent the spectra at $x_h=0$ (half filling), $x_h=1/6$ and $1/3$, respectively. 
Finite size cluster of $N=12$ sites with the open boundary condition is adopted. 
Parameter values are chosen to be $U/t=6$, and $t'/t=1$. 
Bold arrows represent energies of the pump photons adopted in the numerical calculations for the 
real time photoinduced dynamics. 
}
\label{fig:opt_GS}
\end{figure}
In this section, the transient electronic states after the photon pumping are presented. 
The ground states before photon pumping are studied by calculating the spin-gap energy, the holon-binding energy, the optical conductivity spectra and the one-particle excitation spectra and others. 
Most of the results are consistent with the previous calculations in the two-leg ladder Hubbard models.~\cite{noack,noack2,riera}
In Fig.~\ref{fig:opt_GS}, the optical conductivity spectra before pumping at half filling $(x_h=0)$ and away from the half filling are shown. 
The optical gap at half filling is collapsed by hole doping. 
A sharp low-energy peak around $\omega/t=0.5$ which corresponds to the Drude component appears. 
Finite but small energy of the ``Drude" peak is due to the finite size effects in the open boundary condition, 
and is confirmed to decrease with increasing the system size. 
The pump photon energies ($\omega_p$) are tuned at the optical gap energy in the the half filling case, and 
at the energy of the remnant gap in the hole doped metallic state as shown by bold arrows in Fig.~\ref{fig:opt_GS}.
We define the absorbed photon density as $n_{p}(\tau) =(E(\tau)-E_{0})/(N \omega_{p})$ 
where $E(\tau)$ is the total energy at time $\tau$, and $E_{0}=E(\tau \ll 0)$.
Numerical values of the pump photon amplitudes are chosen to satisfy the condition 
$n_{p}(\tau \gg \gamma_{p}) \sim 0.05$.

\subsection{Photoinduced change in carrier dynamics}
\label{sec:kinetics}

\begin{figure}[t]
\includegraphics[width=\columnwidth,clip]{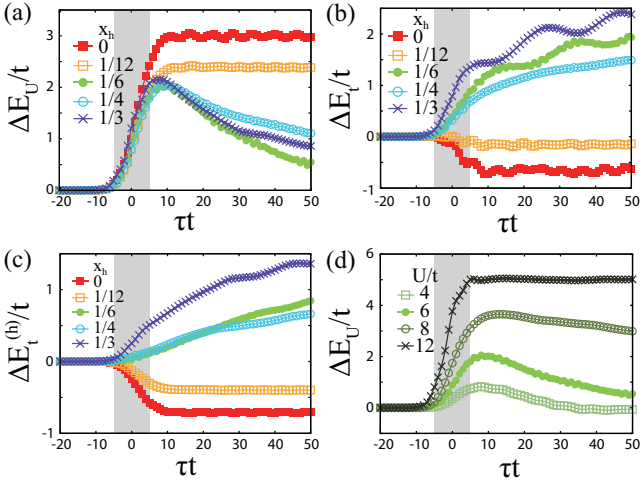}
\caption{
(Color online)
(a)-(c) Increments of several terms of energy for several hole densities.  
The Coulomb-interaction energy $(E_U)$, the kinetic energy $(E_t)$, and the kinetic energy for the low-energy hole motions $(E_t^{(h)})$ are plotted. 
The Coulomb interaction parameter is chosen to be $U/t=6$. 
(d) Time dependences of the Coulomb interaction energy at $x_h=1/6$ for several values of $U$. 
Finite size cluster of $N=12$ sites with the open boundary condition is adopted. 
Parameter value is chosen to be  $t'/t=1$. 
Shaded areas represent the time interval when the pump pulse is introduced. 
}
\label{fig:sta1}
\end{figure}

First, we show the photoinduced changes in the carrier dynamics. 
In order to measure the low-energy dynamics of carriers, the projected kinetic energy expectations are introduced as~\cite{yokoyama} 
\begin{align}
E_{t}^{(h)}=-t\sum_{\langle ij \rangle \alpha \sigma} \langle P_{ij \alpha {\bar \sigma}}^{(h)}  c_{i \alpha \sigma}^\dagger c_{j \alpha \sigma} P_{ij \alpha {\bar \sigma}}^{(h)}+H.c. \rangle , 
\label{eq:kine}
\end{align}
and 
\begin{align}
E_{t}^{(d)}=-t\sum_{\langle ij \rangle \alpha \sigma} 
\langle P_{ij \alpha {\bar \sigma}}^{(d)}  c_{i \alpha \sigma}^\dagger c_{j \alpha \sigma} 
P_{ij \alpha {\bar \sigma}}^{(d)} +H.c. \rangle , 
\label{eq:kine1}
\end{align}
where ${\bar \sigma}=\uparrow (\downarrow)$ for $\sigma=\downarrow (\uparrow)$. 
We define the projection operators as  
$
P_{ij \alpha \sigma}^{(h)}=(1-n_{i \alpha \sigma})(1-n_{j \alpha \sigma}) 
$
and 
$
P_{ij \alpha \sigma}^{(d)}=n_{i \alpha \sigma}n_{j \alpha \sigma}  
$
which project onto the states where both the $i$ and $j$ sites are unoccupied and occupied by electrons with spin $\sigma$, respectively. 
%
The projected kinetic energies $E_t^{(h)}$ and $E_t^{(d)}$, respectively, measure the kinetic energies of holes and doublons along the leg, where the number of the double occupancies are not changed, 
for examples, 
$|(\uparrow )_i \rangle \rightarrow |(\uparrow)_j \rangle$ and 
$|(\uparrow \downarrow )_i  (\downarrow)_j \rangle  \rightarrow |(\downarrow)_i (\uparrow \downarrow )_j\rangle$.

In Fig.~\ref{fig:sta1}, several components of the energies are plotted as functions of time. 
The Coulomb interaction energy ($E_U$) and the kinetic energy ($E_t$), respectively, are defined as the expectation values of the third and first terms in Eq.~(\ref{eq:ham}). 
We define the energy differences as $\Delta E_U=E_U(\tau)-E_U(\tau \ll 0)$ and others. 
As shown in Fig.~\ref{fig:sta1}(a), $E_U$ increases by photon pumping in all values of $x_h$, which implies that the number of the doublely occupied sites increases.
Time dependences of $E_U$ after photon pumping strongly depend on $x_h$: large $E_U$ is maintained at half filling, while it gradually decreases with time in hole doped cases. 
The reductions of $E_U$ imply recombinations of the photo-generated doublons and holons, and the excess Coulomb-interaction energy due to the photon pumping is transfered into the kinetic energy. 
At vicinity of the half filling, there are no channels through which the large excess Coulomb-interaction energies are released. This is shown clearly in Fig.~\ref{fig:sta1}(d) where $E_U$ at $x_h=1/6$ is plotted for several values of $U$.
Reduction of $E_U$ is observed in the cases of small $U$. 
The life times of the photo-generated doublons and holons are prolonged by increasing $U$, as suggested to be $\sim e^{\alpha U}$ with a positive constant $\alpha$ in Ref.~\onlinecite{strohmaier}. 
The kinetic energies, $E_t$ and $E_t^{(h)}$, respectively, are presented in Figs.~\ref{fig:sta1}(b) and (c).  
We confirm that results of $E_t^{(h)}+E_t^{(d)}$ (not shown) show similar behaviors to $E_{t}^{(h)}$.  
At half filling, both $E_t$ and $E_t^{(h)}$ decrease, implying increment of the carrier motions by photon pumping.  
Opposite changes are seen in the hole doped cases where both $E_t$ and $E_t^{(h)}$ increase. That is, the carrier motions are suppressed by photon pumping. 

\begin{figure}[t]
\includegraphics[width=\columnwidth,clip]{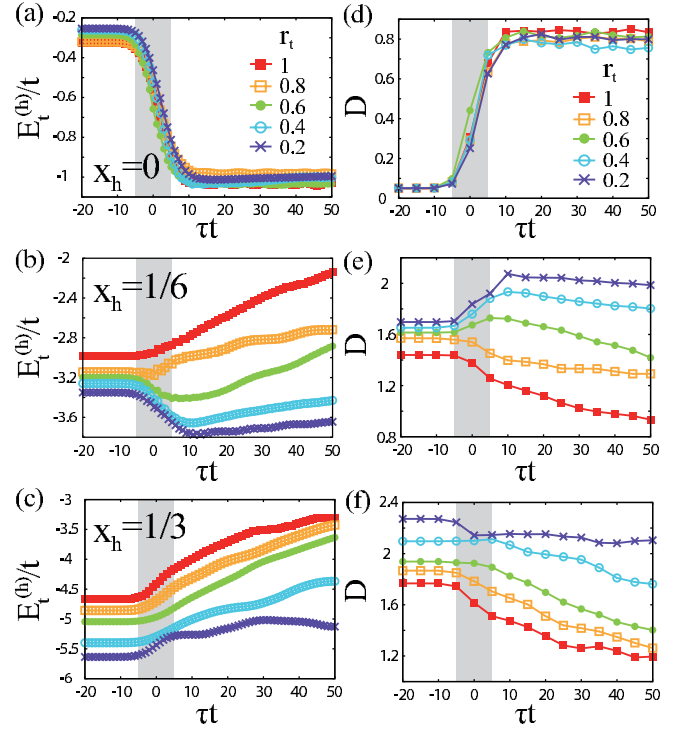}
\caption{
(Color online)
(a)-(c) Kinetic energies of the low-energy hole motions for several values of $t'/t$. 
(d)-(f) Low energy weights of the optical conductivity spectra for several values of $t'/t$. 
Hole densities are chosen to be $x_h=0$ in (a) and (d), $1/6$ in (b) and (e), and $1/3$ in (c) and (f). 
Finite size cluster of $N=12$ sites with the open boundary conditions is adopted. 
Parameter value is chosen to be $U/t=6$. 
Shaded areas represent the time interval when the pump pulse is introduced. 
}
\label{fig:aniso}
\end{figure}
Let us focus on the photoinduced dynamics of the low-energy carrier motion. 
In Figs.~\ref{fig:aniso}(a)-(c), the transient kinetic energies for the low-energy hole motion, $E_t^{(h)}$, are presented for several values of $r_t=t'/t$. 
At half filling (Fig.~\ref{fig:aniso}(a)), reductions of $E_t^{(h)}$ by the photon pumping occur commonly in all values of $t'/t$.  Remarkable $t'/t$ dependences are shown in the hole doped cases of $x_h=1/6$ [see Fig.~\ref{fig:aniso}(b)]. 
Monotonic reductions of $|E_t^{(h)}|$ after photon pumping 
are observed around $r_t=1$. 
In the cases of the weakly coupled two chains corresponding to small $r_t$, 
$|E_t^{(h)}|$ increases by photon pumping as similar to the results at half filling. 
The reductions of $|E_t^{(h)}|$ are  more pronounced in $x_h=1/3$ as shown in Fig.~\ref{fig:aniso}(c). 
Therefore, the two-leg ladder lattice plays an essential role for the suppression of the low-energy carrier dynamics by photon pumping. 
Qualitatively similar behaviors of $E_t^{(h)}$ are confirmed in the calculations, in which 
the cluster sizes are $N=5 \times 2$ to $7 \times 2$ with the periodic and open boundary conditions, and $U/t$ is chosen to be 6 and 8. 

\begin{figure}[t]
\includegraphics[width=0.8\columnwidth,clip]{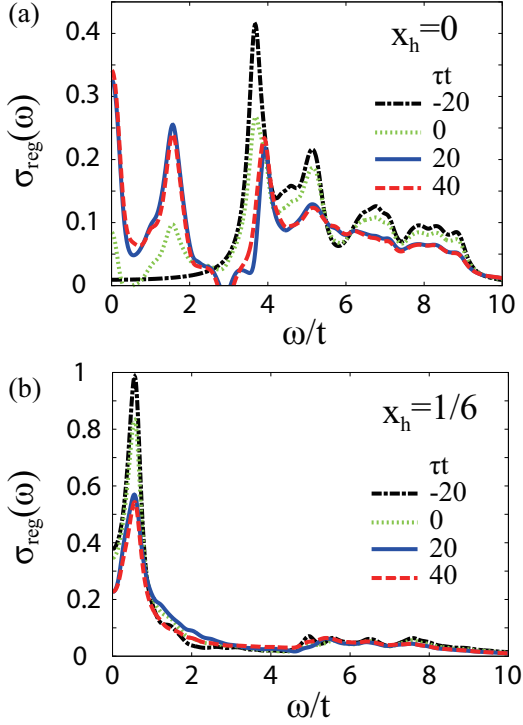}
\caption{
(Color online)
Optical conductivity spectra for several times at (a) $x_h=0$ and (b) $1/6$ . 
Finite size cluster of $N=12$ sites with the open boundary condition is adopted. 
Parameter values are chosen to be $U/t=6$ and $t'/t=1$.
}
\label{fig:opt1}
\end{figure}
\begin{figure}[t]
\includegraphics[width=0.8\columnwidth,clip]{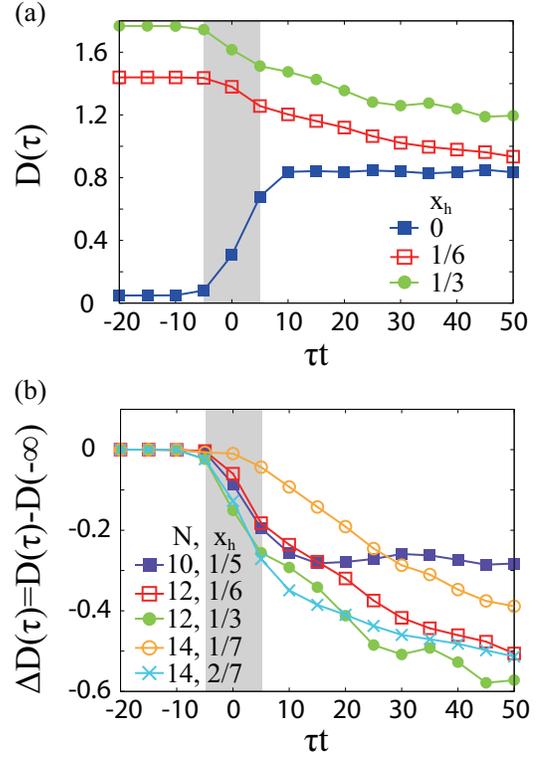}
\caption{
(Color online)
(a) Time dependences of the low-energy weights of the optical conductivity spectra $(D)$ for several hole densities. 
Finite size cluster of $N=12$ sites with the open boundary condition is adopted. 
(b) Changes in the low-energy weights of the optical conductivity spectra for several cluster sizes and hole  densities in the hole-doped cases. 
We chose $(N, x_h)=(10, 1/5)$, $(12, 1/6)$, $(12, 1/3)$, $(14, 1/7)$, and $(14, 2/7)$, respectively. 
Parameter values are chosen to be $U/t=6$, $t'/t=1$ and $\omega_c/t=2$. 
Shaded areas represent the time interval when the pump pulse is introduced. 
}
\label{fig:dt}
\end{figure}
Carrier dynamics after photon pumping are also examined by calculating the transient excitation spectra. 
The transient optical responses are simulated by using the formula based on the linear response theory where the wave functions at time $\tau$ are used. 
This has been utilized widely to examine the transient electronic structures in correlated electron systems,~\cite{nagaosa,kanamori} and its validity was checked in Ref.~\onlinecite{ohara}. 
The regular part of the optical conductivity spectra is given by 
\begin{align}
\sigma_{reg}(\omega)=-\frac{1}{N \omega} {\rm Im}\chi(\omega), 
\end{align}
with  
\begin{align}
\chi(\omega)=- \sum_{m n}
\biggl ( & \frac{\langle \Psi (\tau ) | \phi_m \rangle    \langle \phi_m | j | \phi_n \rangle \langle \phi_n |j | \Psi(\tau) \rangle}{\omega-\varepsilon_m+\varepsilon_n+i \eta}
\nonumber \\
- 
&\frac{\langle \Psi(\tau) |j| \phi_n \rangle    \langle \phi_n | j | \phi_m \rangle \langle \phi_m | \Psi(\tau) \rangle}{\omega-\varepsilon_n+\varepsilon_m+i \eta} \biggr ) , 
\label{eq:chi}
\end{align}
where we introduce the current operator along the leg defined by 
$j=i\sum_{i \alpha \sigma}c_{i \alpha \sigma}^\dagger c_{i+1 \alpha \sigma}+H.c.$, and a small  positive constant $\eta$. 
The calculated optical conductivity spectra at $x_h=0$ and $1/6$ 
are shown in Figs.~\ref{fig:opt1}(a) and (b), respectively. 
At half filling, new peaks appear inside of the optical gap by photon pumping and grow with time. 
This is consistent with the results of $E_t^{(h)}$ shown in Fig.~\ref{fig:sta1}(b). 
This change in the optical conductivity spectra implies a transition from a Mott insulator to a metallic state. 
\begin{figure*}[t]
\includegraphics[width=\textwidth,clip]{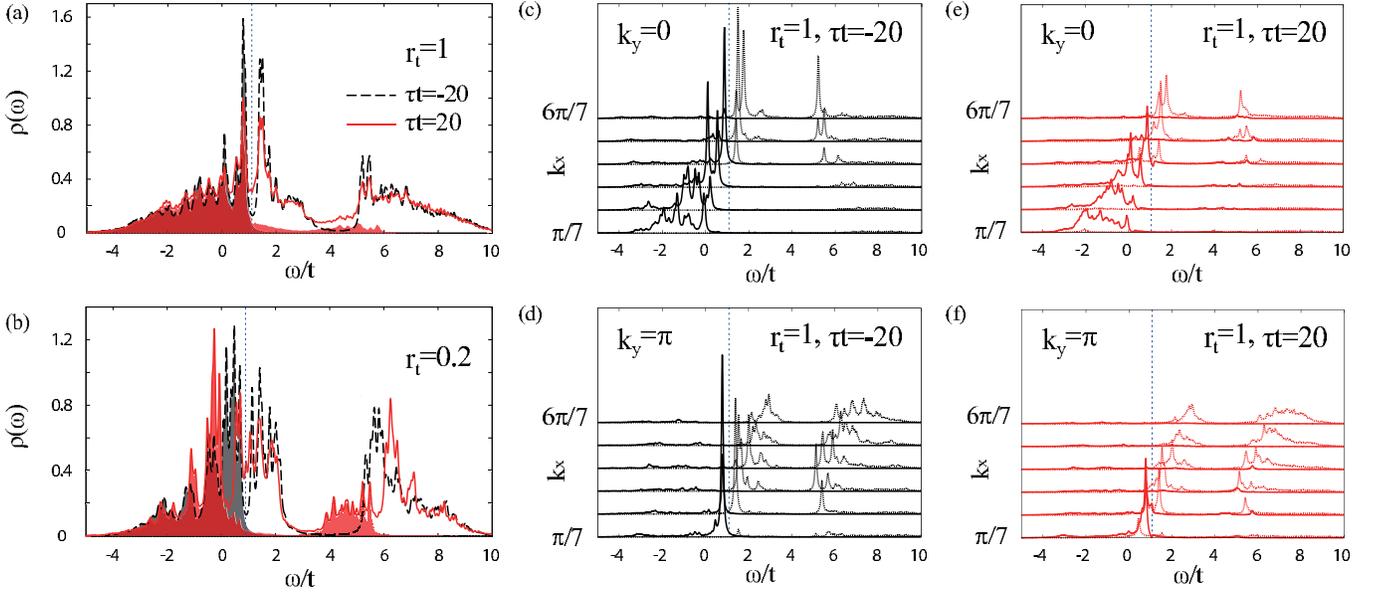}
\caption{
(Color online)
(a) One particle DOS at $r_t=1$ and (b) those at $r_t=0.2$ before $(\tau t=-20)$ and after $(\tau t =20)$ pumping. 
Dashed and solid lines represent the total DOS $[\rho^<(\omega)+\rho^>(\omega)]$ before and after pumping, respectively, and black and red  shaded areas correspond to the electron parts of DOS $[\rho^<(\omega)]$ before and after pumping, respectively. 
Dotted lines denote the Fermi level before pumping. 
(c) and (d) One particle excitation spectra $A^>({\bm k},\omega)$ (solid lines) and $A^<({\bm k}, \omega)$ (dotted lines) before pumping,
and (e) and (f) those after pumping in the case of $r_t=1$. 
Finite size cluster of  $N=12$ sites with the open boundary condition is adopted. 
Parameter value is chosen to be $U/t=6$.
}
\label{fig:dos}
\end{figure*}
On the other hand, in the hole-doped case, a sharp low-energy peak observed before pumping is diminished  just after the photon pumping, and its intensity decreases with time.  
Weak change is seen in high-energy excitation spectra, that are identified as remnants of the Mott gap excitations. 
We note that negative values around $\omega/t=3$ in Fig.~\ref{fig:opt1}(a) is attributable to be due to the optical emission from the photoexcited states. 

In order to examine photoinduced changes in the low-energy spectral weight, we define the integrated spectral weight defined by~\cite{wagner, shimizu}
\begin{align}
D=-\frac{\pi E_{t}}{N}-2 \int_{\omega_c}^{\infty} d \omega \sigma_{reg}(\omega) , 
\end{align}
in the calculations with the open boundary condition, where $\omega_c$ is a cut off energy. 
In Fig.~\ref{fig:dt}(a), the low-energy integrated spectral weights are shown for several hole densities. 
We chose $\omega_c/t=2$, and do not observe any qualitative differences from the results with $1 \lesssim \omega_c/t \lesssim 3$. 
Increase of $D$ from zero at half filling  implies an appearance of the photo-generated metallic-like carrier motions. 
Opposite behaviors are observed in $x_h=1/6$ and $1/3$; $D$ is redued after photon pumping. 
Low energy intensities continuously decrease at least up to $\tau=50/t$. 
These characteristics in the hole doped cases are widely observed in several size clusters and hole densities as shown in Fig.~\ref{fig:dt}(b). 
We confirmed that the reduction of $D$ is monotonically increased with the pump pulse amplitude $A_p$, 
and is qualitatively insensitive to the parameter values of the photon energy between $4.9 \lesssim \omega_p/t \lesssim 7.5$ in $(N, x_h)=(12,1/6)$. 
The low-energy spectral weights for several $x_h$ and $r_t=t'/t$ are summarized in Figs.~\ref{fig:aniso}(d)-(f). The results are qualitatively similar to the kinetic energies for the low-energy hole motion 
shown in Figs.~\ref{fig:aniso}(a)-(c). 

Photoinduced changes in the electronic structures are directly observed by calculating the one-particle excitation spectra. We calculate the transient electronic density of states (DOS) 
where the wave function at time $\tau$ is used.
This is given by 
\begin{align}
\rho^{\gtrless}(\omega)=\frac{1}{N}\sum_{\bm k} A^{\gtrless}({\bm k}, \omega) , 
\end{align}
where $\rho^> (\omega)$ and  $\rho^< (\omega)$, respectively, are the electron and hole parts, and are 
given by the one-particle excitation spectra defined by 
\begin{align}
A^{<}({\bm k}, \omega)&=-\frac{1}{\pi} {\rm Im} 
\sum_{m n }\sum_{\alpha \sigma}  \nonumber \\
& \times 
 \frac{\langle \Psi (\tau ) | \phi_m \rangle    
\langle \phi_m |c_{{\bm k} \alpha \sigma} | \phi_n \rangle 
\langle \phi_n |c_{{\bm k} \alpha \sigma}^\dagger| \Psi(\tau) \rangle}
{\omega-\varepsilon_n+\varepsilon_m+i \eta} ,
\label{eq:chi1}
\end{align}
and 
\begin{align}
A^{>}({\bm k}, \omega)&=-\frac{1}{\pi} {\rm Im}
\sum_{m n }\sum_{\alpha \sigma} \nonumber \\
& \times
\frac{\langle \Psi(\tau) |c_{{\bm k} \alpha \sigma}^\dagger| \phi_n \rangle    
\langle \phi_n | c_{{\bm k} \alpha \sigma} | \phi_m \rangle 
\langle \phi_m | \Psi(\tau) \rangle}
{\omega-\varepsilon_m+\varepsilon_n+i \eta} . 
\label{eq:chi2}
\end{align}
Calculated DOS and one-particle excitation spectra are shown in Fig.~\ref{fig:dos}. 
We take that the $x$ and $y$ axes are parallel to the leg and rung directions, respectively. 
In the isotropic ladder lattice (Fig.~\ref{fig:dos}(a)), a sharp peak around the Fermi level (FL) before pumping is attributed to the two quasi-particle bands originating from the bonding $(k_y=0)$ and anti-bonding $(k_y=\pi)$ bands,~\cite{riera} which cut the FL around 
$k_x=0.6\pi$ and $0.3\pi$, respectively [see Figs.~\ref{fig:dos}(c) and (d)]. 
After pumping, as shown in Figs.~\ref{fig:dos}(a), (e) and (f), sharp quasi-particle peaks are smeared out, and are merged into the incoherent parts.  
Remarkable changes in high energy structure are not seen, although electronic and hole parts distribute to high and low energy regions, respectively. 
On the other hand, at $r_t=t'/t=0.2$, i.e. the case of the weakly coupled two chains, sharp peak around FL before pumping remains and shifts toward the low energy after photon pumping. 
In the numerical calculation shown in Fig.~\ref{fig:dos}(b), 
the pump photon amplitude is chosen to a value in which $n_{p}$ is taken to be about $0.08$ in order to see the characteristics clearly. 
The differences between the results in $r_t=1$ and $0.2$ shown above are consistent with 
the $r_t$ dependence of the low-energy kinetic energies as well as the low-energy weight of the optical conductivity spectra shown in Fig.~\ref{fig:aniso}. 

\begin{figure}[t]

\includegraphics[width=\columnwidth,clip]{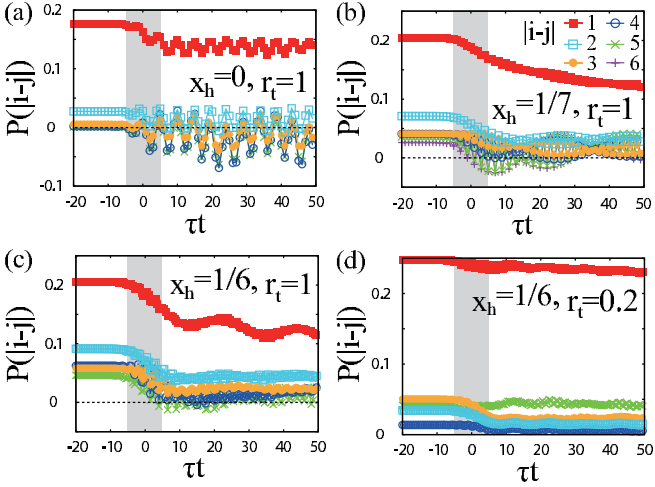}
\caption{
(Color online)
Time dependences of the pair-field correlation functions for several distances $|i-j|$. 
Hole densities are  (a) $x_h=0$, (b) $1/7$ and (c) $1/6$.
Finite size clusters of $N=12$ and $14$ sites with the open boundary condition are adopted.  
Parameter values are chosen to be $U/t=6$ and $t'/t=1$.
Results at $x_h=1/6$ and $r_t=0.2$ are shown in (d). 
Shaded area represents the time intervals when the pump pulses are introduced. 
}
\label{fig:pair1}
\end{figure}
Finally, we show the paring properties of charge carriers and the photon pumping effect.~\cite{noack,noack2,dagotto} 
We introduce the pair-field correlation function between sites $i$ and $j$ defined by 
\begin{align}
P(|i-j|)=\langle \Psi(\tau) |\left ( \Delta_j^\dagger \Delta_i + H.c. \right ) |\Psi(\tau) \rangle , 
\end{align}
where the $d_{x^2-y^2}$-wave pair-field operator is given by 
\begin{align}
\Delta_i=c_{i 1\downarrow}c_{i 1 \uparrow}-c_{i 2 \uparrow}c_{i 1 \downarrow} . 
\end{align}
This function measures the correlation between the carrier pairs  created at sites $j$ and annihilated at $i$. 
As shown by the previous calculations,~\cite{noack,noack2,dagotto} the pair correlations in the equilibrium  states at zero temperature are
damped within few sites at half filling, and are long ranged in the doped cases. 
At half filling [see Fig.~\ref{fig:pair1}(a)],  while the long-range correlation is reduced a little after pumping, 
the short-range correlations are changed to be oscillating with large amplitude.  
Away from the half filling [see Figs~\ref{fig:pair1}(b) and (c)], 
the correlations for all distances monotonically decrease by photon pumping, 
while the oscillatory behaviors observed at $x_h=0$ are weak.  
The reductions in $P(|i-j|)$ are not pronounced in the case of small $r_t=t'/t$ as shown in Figs.~\ref{fig:pair1}(d). 
These results are summarized that i) coherent oscillations of the carrier pairs are induced by the photon pumping at half filling, 
and ii) the pair correlation becomes a short ranged by the photon pumping at hole doped cases. 
Based on the results, the optical control of the carrier pair coherence will be demonstrated in Sect.~\ref{sec:double}.

\subsection{Double pulse pumping}
\label{sec:double}

In this section, we demonstrate the electronic state changes by the double pulse pumping, where the two photon pulses with a time interval are introduced in the two-leg ladder Hubbard model. 
The vector potential for the double-pulse pumping is given by 
\begin{align}
A_{p}(\tau)&=A_1 e^{-\tau/(2\gamma_1^2)} \cos \omega_{1} \tau 
\nonumber \\
&+ 
 A_2 e^{-(\tau-\tau_d) /(2\gamma_2^2)} \cos \omega_{2} \tau , 
\label{eq:vp2}
\end{align}
where $\tau_d$ is the time interval between the first and second pulses, 
$A_{1(2)}$, $\omega_{1(2)}$, and $\gamma_{1(2)}$ are amplitude, frequency and damping factor 
for the first (second) pulse, respectively. 
In  the following calculations, we take 
$\gamma_1 = \gamma_2 = 1/t$. 
Photon energies in  both the first and second pulses are chosen to be the optical gap energy before the first pumping as $\omega_1 = \omega_2 = 4t$.
 
\begin{figure}[t]
\includegraphics[width=0.8\columnwidth,clip]{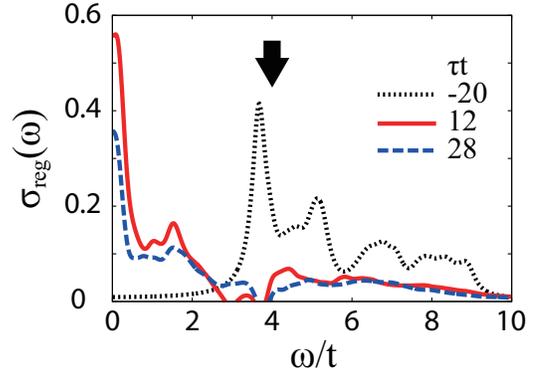}
\caption{
(Color online)
Optical conductivity spectra in the double pulse pumping at half filling ($x_h=0$). 
Dotted, solid and dashed lines represent the spectra before pumping $(\tau t=-20)$, between the first and second pulse pumpings $(\tau t=12)$, and after the second pulse pumping $(\tau t= 28)$, respectively. 
Finite size cluster of  $N=12$ sites with the open boundary condition is adopted. 
Parameter values are chosen to be $U/t=6$, $t'/t=1$, $A_1=A_2=0.6$, $\omega_1=\omega_2=4$, and $\tau_d=15$. Bold arrow represents the energy of the first and second photon pulses. 
}
\label{fig:double_opt}
\end{figure}
The optical conductivity spectra at half filling before the first pulse, between the first and second pulses, and after the 2nd pulse are shown in Fig.~\ref{fig:double_opt}.
By introduction the first pump pulse, the low-energy spectral weight appears inside the optical gap. After the second pump pulse, reduction of the low-energy spectral weight occurs. 
In brief, this is a sequential change of the electronic structures as (a Mott insulator) $\rightarrow$ (a metallic state)  $\rightarrow$ (a suppression of the metallic state). 
In this sense, the photo-doped carriers by the first pump pulse play similar roles with the chemically doped carriers. 

\begin{figure}[t]
\includegraphics[width=\columnwidth,clip]{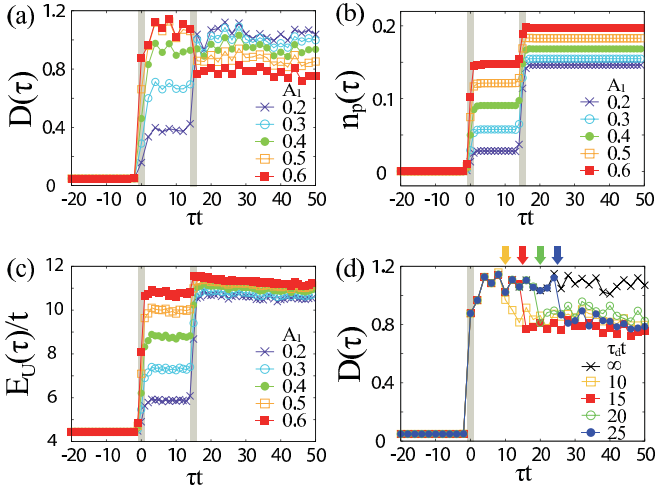}
\caption{
(Color online)
The calculated results in the double pulse pumping at half filling. 
(a) Low energy weights of the optical conductivity spectra, (b) the absorbed photon density and (c) the Coulomb interaction energy. 
The first pulse amplitudes are varied as $A_1=0.2-0.6$ and the second pulse amplitude is fixed to be $A_2=0.6$.  
(d) Low energy weights of the optical conductivity spectra for several values of $\tau_d$.
The first and second pulse amplitudes are fixed to be $A_1= A_2=0.6$. 
Finite size cluster of  $N=12$ sites with the open boundary condition is adopted. 
Parameter values are chosen to be $U/t=6$, and $t'/t=1$. 
Shaded areas represent the time intervals when the two pump pulses are introduced. 
}
\label{fig:double_d}
\end{figure}
We calculate the low-energy weights of the optical conductivity spectra, $D$, as functions of time 
and show the results in Fig.~\ref{fig:double_d}(a), in which $A_1$ are varied and $A_2$ is fixed. 
The spectral weight induced by the first pulse increases with increasing $A_1$. 
On the other hand, the second pulse brings about non-monotonic changes in $D$; when $A_1$ is small (large), the second pulse increases (decreases) $D$. 
The absorbed photon density and the Coulomb interaction energy are shown in Figs.~\ref{fig:double_d}(b) and (c), respectively. 
Not only the first pulse pumping but also the second pulse pumping induces increases both of $n_p$ and $E_U$, in contrast to the results of $D$ shown in Fig.~\ref{fig:double_d}(a). 
This implies that even in the case for the large $A_1$, 
the second pumping realizes a higher energy excited state than the state before the second pumping, but suppresses the low energy carrier motion.
In Fig.~\ref{fig:double_d}(d), we show the low energy spectral weight 
in the case of the large $A_1$ where the time interval between the first and second pulses is varied.
The characteristic reduction of $D$ after the second pumping is commonly observed in all cases of $\tau_d$. 
Thus, the observed phenomena are not due to the interference effects between the first and second pulse excitations.

\begin{figure}[t]
\includegraphics[width=\columnwidth,clip]{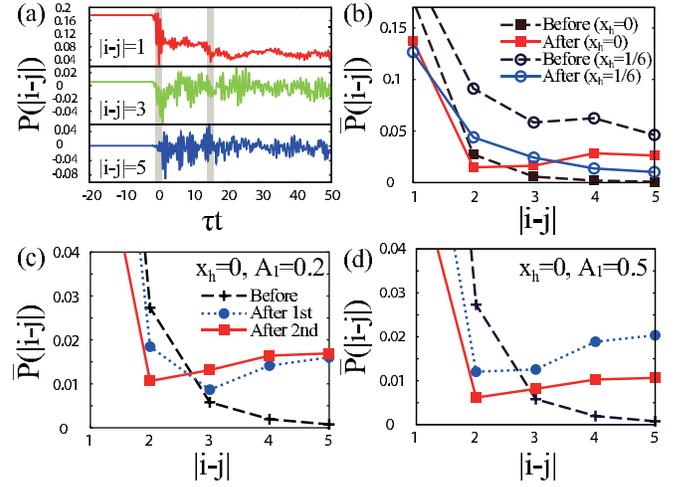}
\caption{
(Color online)
(a) Time dependences of the pair-field correlation functions in the double-pulse pumping at half filling. Upper, middle and lower panels show $P(|i-j|)$ for $|i-j|=1$, $3$, and $5$, respectively. 
(b) Time averaged pair-field correlation functions in the single-pulse pumping at half filling ($x_h=0$) and at  hole doped  case $(x_h=1/6)$. 
Dashed and solid lines represent ${\bar P}(|i-j|)$ before and after photon pumpings, respectively. 
(c) and (d)
Time averaged pair-field correlation functions in the double-pulse pumping at half filling. 
The first pulse amplitudes are taken to be $A_1=0.2$ in (c) and $A_1=0.5$ in (a) and (d). 
The second pulse amplitude is $A_2=0.6$ in (a), (c) and (d). 
Dashed, dotted, and solid lines represent ${\bar P}(|i-j|)$ before the first pulse, between the first and second pulses, 
and after the second pulse, respectively. 
Finite size cluster of the $N=12$ sites with the open boundary condition is adopted. 
Parameter values are chosen to be $U/t=6$, and $t'/t=1$. 
}
\label{fig:double_pair}
\end{figure}
The pairing characteristics are examined under the double pulse pumping. 
Time dependences of the pair-field correlation functions, $P(|i-j|)$, for $|i-j|=3-5$ at half filling are plotted in Fig.~\ref{fig:double_pair}(a).
As shown previously, $P(|i-j|)$ before the first pulse is of short ranged; $P(|i-j|>3)$ is less than 5$\%$ of $P(|i-j|=1)$. 
After the first pulse pumping, coherent oscillations appear in $P(|i-j|)$ for all $|i-j|$ and their maximum amplitudes of the oscillations are larger than 10$\%$ of $P(|i-j|=1)$ before the first pumping. 
After the second pulse pumping, suppressions of the oscillation amplitudes are remarkably seen in $P(|i-j|)$ for  $|i-j|=4$ and $5$.   

We analyze this characteristic change in the pair-field correlation function by introducing an averaged absolute value of $P(|i-j|)$ 
after the first and second pulse pumpings defined by 
\begin{align}
{\bar P}(|i-j|)=\frac{1}{\Delta t} \int_{t_0}^{t_0+\Delta t} dt | P(|i-j|) | , 
\end{align}
where the parameters $(t_0, \Delta t)$ are chosen to be $(0, 15)$ in the case after the first pulse, and $(15, 35)$ in the case after the second pulse. 
Results are shown in Figs.~\ref{fig:double_pair}(c) and (d) for the large and small $A_1$s, respectively. 
For comparison, we show in Fig.~\ref{fig:double_pair}(b) the calculated results in the single pulse pumping case at  $x_h = 0$ and $1/6$ where we take $(t_0, \Delta t)=(15, 35)$.
After the first pulse pumping 
(see the blue dotted lines in Figs.~\ref{fig:double_pair}(c) and (d)), ${\bar P}(|i-j|)$ for $|i-j| \ge 3$ increases and the correlation becomes long ranged. 
These results are similar to that in the single pumping case at $x_h=0$ shown by the red line in Fig.~\ref{fig:double_pair}(b). 
The stronger the first pump amplitudes are, the larger the change in ${\bar P}(|i-j|)$. 
Results after the second pulse pumping are qualitatively different between the cases with the large and small $A_1$s; 
${\bar P}(|i-j|)$ with $|i-j| \ge 3$ increase in the small $A_1$ case, but decrease in the large $A_1$. 
In short, the second pulse promotes the pair correlation in the small $A_1$ case, and suppresses the pair coherence in the large $A_1$ case. 
This change in ${\bar P}(|i-j|)$ by the second pulse pumping in the arge $A_1$ case is similar to that 
by the single pumping in the hole doped case as shown in Fig.~\ref{fig:double_pair}(b), 
where the pair correlation becomes shortened by the pumping. 
These results suggest a similarity of the chemically doped and photo-doped hole carriers.

\subsection{Discussion and Summary}
\label{sec:summary}

In this section, we discuss relations of the present theoretical calculations with the recent optical pump-pulse experiments in the ladder cuprates. 
The photoinduced electronic state transition was observed by the time resolved optical spectroscopy in insulating Sr$_{14}$Cu$_{24}$O$_{41}$ in which the nominal valence of Cu ion is +2.25 and 0.25 hole per Cu ion exists.~\cite{fukaya} 
The insulating nature confirmed from the transport and optical measurements is attributable to the 
hole carrier localization in the charge density wave state. 
The pump pulse energy was tuned around the insulating gap energy (1.58eV). 
After the photon pumping, the Drude like metallic state appears with 1ps and is maintained for more than 50ps. 
The photoinduced metallic reflectivity increases with increasing the pump photon fluence. 
These results are attributable to the photo doping carriers and collapse of the carrier localization. 
Similar photoinduced transitions from an insulating state to a metallic state have been widely confirmed experimentally
in correlated insulating states ~\cite{ogasawara,iwai,okamoto,okamoto1,okamoto2,okamoto3,tajima,kawakami}, such as La$_2$CuO$_4$. 
This photoinduced metallic state is also demonstrated by the present calculations, although the carrier concentration is set to be $x_h$ which is different from that in Sr$_{14}$Cu$_{24}$O$_{41}$.
 As shown in Fig.~\ref{fig:opt1}(a), just after the photon pulse pumping, a low energy spectral weight appears inside the insulating gap, which produces the metallic reflectivity spectra. 
This characteristic photoinduced change of the electronic structure in the insulating state are insensitive to the anisotropy in the transfer integrals ($r_t$) as shown in Figs.~\ref{fig:aniso}(a) and (d),
and might be common properties in a wide class of insulating states realized by the electronic interactions.~\cite{yonemitsu,lu2,iyoda,eckstein} 

The time resolved optical experiments were also performed in the metallic ladder cuprates, 
Sr$_{14-x}$Ca$_x$Cu$_{24}$O$_{41}$ with $x=10$.~\cite{fukaya1,fukaya2} 
The pump photon energy was tuned at 1.58eV corresponding to the charge transfer excitation energy in the ladder plane. 
Just after the photon pumping, the low energy optical reflectivity at 0.5eV shows a reduction, implying a suppression of an initial metallic character.  
The experimental reflectivity shows a slow increasing after around 0.5ps, which is interpreted to be  thermalization through relaxations to the lattice degree of freedom. 
The present theoretical calculations provide a possible interpretation for this experimental observation of the photoinduced suppression of the metallic state. 
We confirm this photoinduced suppression through not only the calculations of the pump-probe spectra, but also the calculations of the low energy component of the kinetic energy, the one-particle excitation spectra and others. 
As shown in Fig.~\ref{fig:aniso}(b)-(f), we identify that the ladder lattice effect plays an essential roles on this suppression, in contrast to a photoinduced metallic state in the Mott insulating state. 

The double pulse pumping reinforces a validity of the present scenario for the photoexcited state in 
metallic ladder cuprates. 
The time resolved spectroscopy under the double pulse pumping was performed experimentally in insulating Sr$_{14}$Cu$_{24}$O$_{41}$.~\cite{fukaya2} 
The first pump pulse fluence was varied and the second pulse fluence was fixed, in the same way with the present theoretical calculation introduced in Sec.~\ref{sec:double}. 
Increases of the reflectivity at 0.5eV were observed monotonically as function of the first pulse fluence. 
On the other hand, the change in the reflectivity by the second pulse pumping qualitatively depends on the first pulse fluence; the reflectivity within 0.5ps after the second pulse pumping increases (decreases) in the cases of the weak (strong) first pulse fluence. 
This tendency of the reflectivity change is well reproduced qualitatively by the present calculations shown in Fig.~\ref{fig:double_opt}. 
In the experiments, over 0.5ps after the second pulse pumping, the reflectivity starts to increase being independent of the second pulse fluence. 
This is attributable to the relaxation effects to the lattice degree of freedom which is not taken into account in the present theoretical calculations. 

In conclusion, we study photoinduced carrier dynamics in a correlated electron system on a two-leg ladder lattice. The ladder Hubbard model is analyzed by utilizing the numerical exact diagonalization method in finite size clusters based on the Lanczos algorithm. 
Through the calculations of the transient low-energy kinetic energy, optical conductivity spectra, one-particle excitation spectra and others, we find that the initial metallic state is suppressed by the photon pulse pumping. 
This is in contrast to the photoinduced metallic state in the Mott insulating state. 
The ladder lattice effect plays an essential role on this photoinduced suppression of the metallic character. 
Calculations of the double pulse pumping and comparison with the corresponding pump-probe experiments 
convince us of the characteristic photoinduced excited states in this correlated electron system on the two-leg ladder lattice. 

\begin{acknowledgments}
We thank  S.~Koshihara, Y. Okimoto, R.~Fukaya , M.~Naka, and J.~Nasu for their helpful discussions.
This work was supported by KAKENHI from the Ministry of Education, Science and Culture of 
Japan (No. 26287070). Some of the numerical calculations were performed using the supercomputing facilities at ISSP, the University of Tokyo. 
\end{acknowledgments}



\end{document}